\documentclass[review,12pt]{elsarticle}
\usepackage[utf8]{inputenc}
\usepackage[T1]{fontenc}
\usepackage[english]{babel}

\usepackage{amsmath,amssymb}

\usepackage{graphicx}
\usepackage{dcolumn}
\usepackage{bm}
\usepackage{xcolor}
\usepackage[normalem]{ulem}

\usepackage[bookmarksnumbered,raiselinks,breaklinks]{hyperref}
\hypersetup{
colorlinks,
citecolor=blue,
filecolor=blue,
linkcolor=blue,
urlcolor=black
}

\usepackage[capitalise,nameinlink]{cleveref}

\usepackage{tcolorbox}
\tcbuselibrary{listings,breakable}

\definecolor{Myorange}{cmyk}{0,0.42,1,0}

\begin{document}
\begin{frontmatter}
\title{Graph-based analysis of inflammatory profiles in New Onset Refractory Status Epilepticus (NORSE)}
\author[inst1]{Linon Denis}
\author[inst1]{Martin Guillemaud}
\author[inst1,inst2]{Vincent Navarro}
\author[inst1,inst2,inst3]{Aurélie Hanin \corref{cor1}}
\cortext[cor1]{Corresponding author}
\ead{aurelie.hanin@icm-institute.org}
\author[inst5]{Mario Chavez}
\address[inst1]{Paris Brain Institute (ICM), Inria Paris, INSERM-U1127, CNRS-UMR7225, Sorbonne University, Pitié-Salpêtrière Hospital, Paris, France}
\address[inst2]{Department of Neurology, Epilepsy Unit, Pitié-Salpêtrière University Hospital, Paris, France}
\address[inst3]{Department of Metabolic Biochemistry, DMU BioGeMH, AP-HP, Sorbonne Université, Pitié-Salpêtrière-Charles Foix University Hospital, Paris, France}
\address[inst5]{CNRS, Pitié-Salpêtrière Hospital, Paris, France}

\begin{abstract}
\textbf{Background and Objectives: }Cryptogenic new-onset refractory status epilepticus (cNORSE) represents one of the most severe forms of status epilepticus, occurring in patients without prior neurological disease, and remaining of unknown aetiology despite extensive diagnostic evaluation. Emerging evidence supports a role for immune dysregulation in cNORSE; however, marked heterogeneity in inflammatory signatures has been reported, complicating the selection of targeted immunotherapies. Therefore, a critical need for tools facilitating the interpretation of cytokine panels exists.\\
\textbf{Methods: }Building on the identification of distinct inflammatory groups of cNORSE patients using a graph clustering approach applied to a cohort of 62 patients with serum profiling of 96 cytokines, we tailored new models to quantify attribution probability to biologically validated clusters. Statistical assessment of the most informative model involved Monte-Carlo simulations and custom-developed parametric tests. Ultimately, we applied our framework to the implementation of a clinician-friendly interface for inflammatory profiling. \\
\textbf{Results: }Our approach enables quick processing of several cytokine profiles, providing the most likely inflammatory cluster, associated attribution probability, and statistical confidence. For longitudinal assessments, the proposed method may also allow tracking the evolution of inflammatory trajectories over time. \\
\textbf{Conclusion: }Systematic statistical characterization of the inflammatory heterogeneity in cNORSE requires the development of clinically actionable support tools. Our study offers a framework that may support personalized immunomodulatory strategies in cNORSE patients through clustering-based cytokine profiling.
\end{abstract}

\end{frontmatter}

\section{Introduction}
Status epilepticus (SE) is a life-threatening condition characterized by prolonged epileptic seizures \cite{trinka_definition_2015}. About 25\% of patients do not respond to conventional anti-seizure medications and are classified as having refractory SE (RSE) \cite{hirsch_proposed_2018}. Among them, around 20\% have no prior neurological disease and no identifiable cause within the first few days of presentation, fulfilling criteria for new onset refractory status epilepticus (NORSE) \cite{hirsch_proposed_2018}. In more than 50\% of NORSE cases, no aetiology is identified despite extensive diagnostic evaluation, defining cryptogenic NORSE (cNORSE) \cite{perin_new_2022}. Mortality remains high, around 10\% of children, and even higher (16-27\%) in adults \cite{carson_severe_2024, tharmaraja_etiology_2023}. Moreover, survivors frequently develop long-term neurological sequelae and chronic epilepsy \cite{perin_new_2022}, \cite{carson_severe_2024}, \cite{tharmaraja_etiology_2023}.

In 2022, international expert consensus guidelines recommended initiation of first-line immunotherapy within three days in patients with NORSE (i.e., corticosteroids, intravenous immunoglobulins, or plasmapheresis); followed by second-line immunotherapy within seven days for cNORSE patients \cite{wickstrom_international_2022}. Several second-line immunotherapies are currently used. Some target innate immune pathways, such as anakinra, as an interleukin 1 receptor antagonist, or tocilizumab, a monoclonal antibody blocking interleukin 6 receptor. Others target adaptive immunity, including rituximab, an anti-CD20 monoclonal antibody depleting B cells, or cyclophosphamide, which modulates T-cell responses \cite{hanin_second-line_2024}. However, no evidence-based strategy currently exists to guide personalized therapeutic decisions, and responses to immunotherapy remain highly variable, emphasizing the need for a better characterization of immune dysregulations in cNORSE.

Previous studies have demonstrated elevated levels of innate immunity-related cytokines (CXCL8, CCL2, CCL3) in 51 patients with cNORSE compared to 47 patients with RSE of known aetiology. Higher serum and cerebrospinal fluid cytokine levels were associated with poorer short and long-term outcomes \cite{hanin_cytokines_2023}.  However, cytokine profiles were heterogeneous across patients, suggesting the existence of distinct inflammatory subtypes. In 2025, we expanded this work to identify biological subgroups among patients with cNORSE \cite{guillemaud_identification_2025}. 

Using a 96-cytokine serum panel in 62 patients with cNORSE, we applied a graph-based clustering approach that identified three distinct inflammatory endotypes \cite{guillemaud_identification_2025}. Cluster A comprised 13 patients with no specific inflammatory marker. Cluster B included patients characterized by predominant activation of acute innate responses, encompassing pathways involved in leukocyte recruitment and migration as well as activation of innate inflammatory signaling mediators. Cluster C consisted of patients exhibiting dysregulation of adaptive immune pathways, suggesting a potential therapeutic benefit from agents targeting adaptive immunity, such as rituximab, cyclophosphamide, or JAK-STAT inhibitors. Patients assigned to cluster B were more likely to develop chronic post-NORSE epilepsy and experienced the poorest clinical outcomes. These findings suggest that, within this subgroup, therapeutic strategies targeting innate immune mechanisms, including anakinra, tocilizumab, anti-TNF$\alpha$ agents, or intrathecal corticosteroids, may represent the most appropriate treatment approach.

Similar cytokine-based profiling approaches have previously been applied to asthma groups \cite{seys_cluster_2017}, obstructive sleep apnoea through t-SNE and UMAP projections \cite{wang_patients_2021}, prognosis in non-small-cell-lung cancer \cite{barrera_cytokine_2015}, and a wide range of other disorders \cite{turner_identification_2018, bai_hierarchical_2021, tomassen_inflammatory_2016, morse_patterns_2019}. However, t-SNE and UMAP are primarily visualization and exploratory tools rather than inferential clustering methods. Both approaches may alter the representation of distances and relationships present in the original high-dimensional space, and the resulting embeddings can be influenced by algorithm-specific parameters and random initialization. Consequently, the clinical application of these clustering approaches remains limited by technical challenges, particularly the interpretation of extended cytokine panels in newly evaluated patients. Few of these clustering approaches proposed graph-based models.

Here, we aimed to demonstrate that graph-based statistical models can be implemented to characterise inflammatory profiles in newly evaluated patients. We developed a probabilistic approach to assign new patients to previously identified clusters. Additionally, we introduced a similarity metric based on dimensionally reduced vectorized cytokine representations of inflammatory profiles, paving the way for longitudinal cluster-tracking of inflammatory profile evolution over time. Ultimately, we applied our framework to the design of an innovative clinician-friendly interface to bridge the gap between prior research findings and their clinical applicability. 

\section{Methods}
\subsection{Addition of a new patient to the reference graph}
This study builds on the graph-based clustering procedure implemented in \cite{guillemaud_identification_2025}, used to identify characteristic inflammatory profiles in a cohort of 62 reference patients \cite{guillemaud_identification_2025, blondel_fast_2008}. We defined the weighted graph of 62 reference patients as $G=(V,E,W)$, where $V=(V_i),i=1,…,N$ are the reference patients (vertices or nodes) with $N=62$, $E$ is the set of undirected links or edges connecting two patients, and $W$ are the set of weights characterizing the similarity levels between two patients connected by an edge. Let $C$ be the graph partition obtained through the reference patient clustering algorithm originally presented in \cite{guillemaud_identification_2025}, we write $C:={C_1, \dots,C_K}$, where the different $C_k$ communities are node-disjoint subgraphs whose nodes are more intra-connected than inter-connected to the rest of the graph. 

To incorporate a new patient into the graph, we first calculated its representative triple-coordinate vector according to the three biological variables defined through PCA training on the reference cohort of 62 patients. Specifically, the loading vectors (eigenvectors) corresponding to the first principal component (PC1) of each of the three biological groups were derived from the training cohort and subsequently retained. The biological profile of the new patient was then projected onto these predefined PC1 axes, yielding three coordinates corresponding to PC1 of Group 1, PC1 of Group 2, and PC1 of Group 3. This procedure ensured that the new patient was represented within the same reduced-dimensional space as the reference cohort. The resulting coordinate vector was subsequently appended to the reference dataset, allowing integration of the new patient into the graph-based analysis. 

\begin{figure}[h]
\centering
\includegraphics[width=1.00\textwidth]{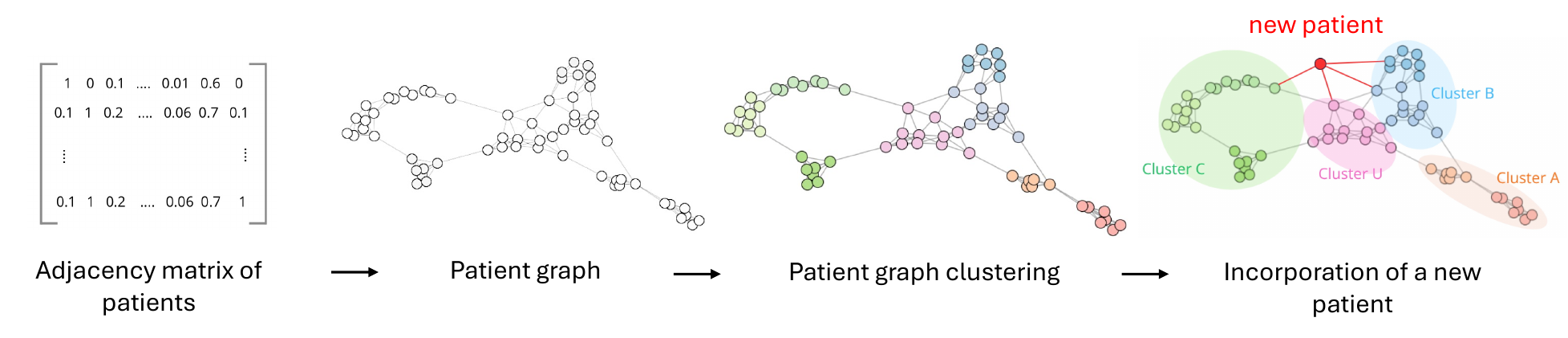}
\caption{Incorporation of a new patient to the reference patient graph. From the adjacency matrix of patients to the graph representation, clustering and incorporation.}
\label{figure1}
\end{figure}

A new patient was therefore incorporated to graph $G$, by connecting its projected node to its $\eta$ closest neighbours, as illustrated in Fig.~\ref{figure1}. Parameter $\eta$ had been set to $\eta=4$  in our previous work \cite{guillemaud_identification_2025}, and yielded coherent results for downstream analyses in this study.

\subsection{Probability-based quantification of cluster assignments for new patients}
To assign clusters to a newly added patient in the extended graph $\tilde{G}=G \cup (V_{N+1},E_{N+1},W_{N+1})$, we tailored a statistical approach where the new patient N+1 is assigned to the closest cluster around according to shortest path (SP) distances on graph $\tilde{G}$. This SP model was compared and selected among two other tailored assignment strategies, namely the Closest Neighbours (CN), and the Euclidean models (EU), which are described in Supplementary Methods. The SP model, relied on the identification of the shortest weighted path $SP(N+1,C_k)$ between the newly added node N+1 and the nodes belonging to each candidate community $C_k$ with $k\in{1,2,…,K}$ (here K=8). As the reference graph $G$ was fully connected, shortest-path distances could be computed for all node pairs. In the event of a disconnected graph, nodes that were not reachable from the new node were assigned an infinite distance $SP(N+1,C_k )=+\infty$.  

For each community $C_k$ a transformed proximity score was computed as:
\begin{equation}
SP_{score}(N+1, C_k) =
\begin{cases}
f(SP(i, C_k)), & \text{if } SP(i, C_k) < +\infty, \\
0, & \text{else.}
\end{cases}
\tag{1}
\end{equation}
where $f$ is a distance-transformation function that can be adjusted to modulate the influence of graph topology on cluster assignment. Several transformation functions were investigated in this study, including : $f_\alpha\left(x\right)={exp}^{-\alpha\ln{\left(x+1\right)}},\alpha\in\mathbb{R}$. The choice of transformation parameter $\alpha$ enabled continuous adjustment of the relative contribution of nodes located in close topological proximity to the new patient versus those located farther away within the graph. Consequently, the model could place varying emphasis on local graph structure or on broader community-wide connectivity patterns when determining the most likely cluster assignment for the new patient.

The attribution probability was then derived from the normalisation of the $SP_{score}$ on the sum of all scores calculated across all possible  communities. For each community $C_k$, the probability of new patient N+1 being attributed to that community (2) was then defined as:
\begin{equation}
P_{N+1,k} = \frac{SP_{score}(N+1, C_k)}{\sum_{j=1}^K SP_{score}(N+1, C_j)}
\tag{2}
\end{equation}
This normalization yields a probabilistic measure of community membership for the newly added node. 

The patient was ultimately attributed to the graph community associated with the highest membership probability $\max_{k \in \{1, \dots, K\}} P_{N+1,k}$. The corresponding biological cluster was then inferred from the predefined mapping between graph communities ($C_k$) and inflammatory clusters. In the present study, this mapping was defined as follows: Cluster A = $\{C_3, C_4\}$, Cluster B = $\{C_1, C_5\}$, Cluster C = $\{C_0, C_2, C_6\}$, and Cluster U = $\{C_7\}$. Cluster-level membership probabilities were obtained by summing the probabilities of the constituent communities. Consequently, the probability of assignment to a given cluster corresponded to the cumulative probability associated with all communities belonging to that cluster. By construction, the resulting cluster probabilities were normalized such that their sum was equal to one.

\subsection{Statistical characterization of the precision of patient attributions}
To evaluate the reliability of patient assignments generated by the shortest-path (SP) probability model, we developed a statistical framework, termed the $\Delta$-test. The objective of this test was to assess whether the assignment of a newly added patient to a given cluster was statistically more likely than assignment to any of the alternative clusters. To establish the null distributions required for inference, we first implemented a Monte Carlo procedure simulating random patient assignments within the graph. This procedure enabled the characterization of the expected probability distributions associated with clusters A, B, C, and U under the null hypothesis of random attribution. The $\Delta$-test subsequently quantified the degree to which the observed assignment probability of a given patient deviated from these null distributions, thereby providing a statistical measure of confidence in the cluster assignment. 

\subsubsection{Monte-Carlo procedure for random simulation}
\begin{figure}[htbp]
\centering
\includegraphics[width=\columnwidth]{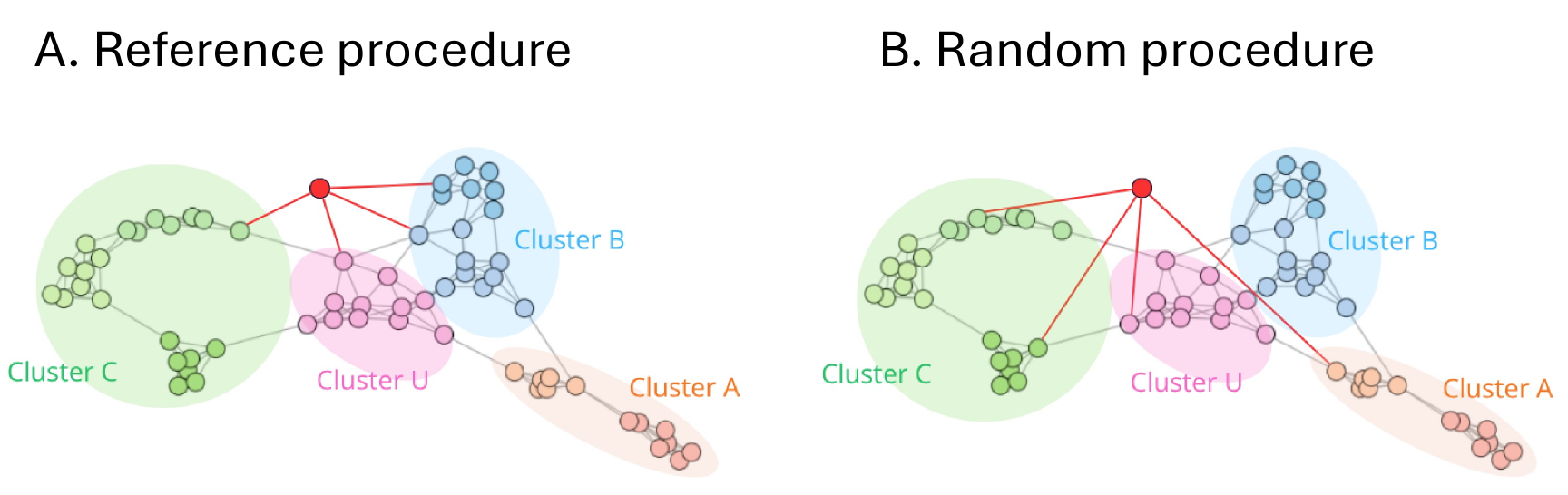}
\caption{The random simulation procedure for adding a new patient node (represented in red) to the reference graph G. \textbf{A.} The reference procedure. \textbf{B.} An example of the random procedure. The graph structure is derived from figures of \cite{guillemaud_identification_2025}. }
\label{figure2}
\end{figure}

The random assignment of a new patient to the existing graph communities was modelled using a Monte Carlo procedure in which the connectivity pattern of the newly added node was randomized. Specifically, rather than connecting the new patient to its nearest neighbours in the graph, edges were established between the new node and randomly selected nodes from the reference graph (see Fig.~\ref{figure2}). Let $x_m=(w_{m,1},w_{m,2},…,w_{m,n})$ denote the connectivity vector of a new patient m where $w_{m,i}$ represents the edge weight between the new node and the reference node $i$. A total of $T=1000$ Monte Carlo iterations were performed. At each iteration $t$, the connectivity pattern of patient $m$ was randomly reassigned while preserving the overall structure of the connectivity vector, thereby generating a randomized patient representaion $x_{m}^{t}$. This procedure yielded a set of $T$ randomized realizations of the original patient. 

At each iteration $t$, a new graph was constructed by incorporating the corresponding randomized patient representation $G^{(t,m)}=G \cup (V^{(t,m)},E^{(t,m)},W^{(t,m)})$. For each simulated graph, the cluster-assignment procedure described above was applied, yielding an attribution of the randomized patient to one of the four clusters, $C_L$ (= A, B, C, or U). The probability of assignment to cluster $C_L$ under the random-connectivity model was then estimated empirically as (3):
\begin{equation}
P^m(C_L) = \frac{1 + t_L}{1 + T}
\tag{3}
\end{equation}
where $t_L$ is the number of Monte Carlo realizations assigned to cluster $C_L$. By Monte-Carlo convergence, these empirical probabilities provided an approximation of the null-hypothesis probabilities $P_0^m (C_L)$ corresponding to the expected cluster-assignment probabilities for patient $m$ under random connectivity. These null distributions subsequently served as the basis for statistical inference in the $\Delta$-test.

\subsubsection{Parametric test implementation: the $\Delta$-test}
The $\Delta$-Test was developed based on the difference $\Delta$ between the highest attribution probability to the most-likely cluster and the second highest attribution probability to a second most-likely cluster. For patient $m$ assigned to cluster $C_L$, we denoted $\Delta$ (4) as:
\begin{equation}
\Delta = P^m(C_L) - \max_{j \neq L} \left( P^m(C_j) \right)
\tag{4}
\end{equation}

Overall, we interpreted $\Delta$ as the assignment precision. Following the simulations (see section 2.3.1), we empirically observed for $M\geq 50$ cases, that the $\Delta$ distribution could be approximated by a normal distribution (see Supplementary Methdos): $\hat{\Delta} \sim N(\mu_0,\sigma_0^2)$ where the parameters $\mu_0$ and $\sigma_0$ denote the empirical estimators of mean and standard deviation, respectively. Therefore, the statistical unilateral $\Delta$ parametric test implemented yielded p-values from the $Z\text{-score}$: $Z = \frac{\hat{\Delta}_{\text{obs}} - \hat{\mu}_0}{\hat{\sigma}_0}$, with $\Delta_{obs}$ being the observed values for $\Delta$. This $\Delta$-test enabled us to assess the contrast superiority in the true cluster attributions we were observing, as compared to the simulated random incorporations of patients to the reference graph $G$. One p-value was generated for each new patient added through this statistical approach. 

\subsection{Similarity measure for longitudinal analyses}
For longitudinal analyses, we quantified temporal changes in inflammatory profiles by measuring the Euclidean distance between biological samples represented in the PCA-derived feature space. Each sample was projected onto the first principal component of the three cytokine-variable groups ($PC_{1,group1}, PC_{1,group2}, PC_{1,group3}$), yielding a three-dimensional representation. Given two samples, $s_1$ and $s_2$ obtained either from different patients or from the same patient at different time points, their biological divergence was defined as $s_{EU}^2 = \| s_1 - s_2 \|^2$. This metric provides a quantitative measure of inflammatory profile evolution, with increasing distances corresponding to greater biological changes over time.

\subsection{Datasets}
The data used in this study included a reference cohort of 62 cNORSE patient samples [9], as well as 50 new cNORSE patient samples (26 new patients, and 24 longitudinal samples from the initial cohort of 62 reference patients), all analysed using the same Eve Technologies’ Human Cytokine 96-plex discovery assay (Canada). The detailed list of the 96 serum markers used in this work is available in the Supplementary section. This study was approved by the Paris Pitié-Salpêtrière Hospital (APHP, COLETTE and Inserm, TIPI) and Yale University (NORSE/FIRES biorepository, IRB \#1511016840 and \#2000031611). Informed consent was obtained from all patients or legally authorized representatives following the Declaration of Helsinki. Control patient data was also used for the purpose of this study, including cases of RSE (not NORSE), chronic epilepsies, encephalitis, and healthy controls. For simplicity, we defined datasets NORSEREF referring to the 62 original patients, NORSENEW referring to the full 50 new patients, and among NORSENEW we defined a subgroup NORSELONGI referring to second-time samples of original patients. Further data detail is provided in the summary Table~\ref{table1}.

\begin{table}[ht]
\centering
\caption{Data summary table (including data notations). Asterisks refer to Ref.~\cite{guillemaud_identification_2025}}
\label{table1}
\resizebox{\textwidth}{!}{%
\begin{tabular}{|l|cc|cc|c|c|c|c|}
\hline
&
\multicolumn{2}{c|}{NORSE} &
\multicolumn{2}{c|}{RSE} &
CHRONIC &
ENCEPHALITIS &
NORSELONGI &
HCONTROL \\
\hline

&
NORSEREF* &
NORSENEW &
\begin{tabular}[c]{@{}c@{}}
From reference\\
article*
\end{tabular} &
\begin{tabular}[c]{@{}c@{}}
New\\
samples
\end{tabular} &
\begin{tabular}[c]{@{}c@{}}
New\\
samples
\end{tabular} &
\begin{tabular}[c]{@{}c@{}}
New\\
samples
\end{tabular} &
\begin{tabular}[c]{@{}c@{}}
New\\
samples
\end{tabular} &
\begin{tabular}[c]{@{}c@{}}
New\\
samples
\end{tabular}
\\
\hline

Patients &
p0--p61 &
p62--p87 &
C62--C106 &
C107 &
ch0--ch49 &
E0--E18 &
L0--L23 &
H0--H17 \\
\hline

Effect size &
62 &
26 &
45 &
1 &
50 &
19 &
24 &
18 \\
\hline
\end{tabular}}
\end{table}

\subsection{The interface \textit{NORSE Profiles}}
\begin{figure*}[t]
\centering
\includegraphics[width=1.00\textwidth]{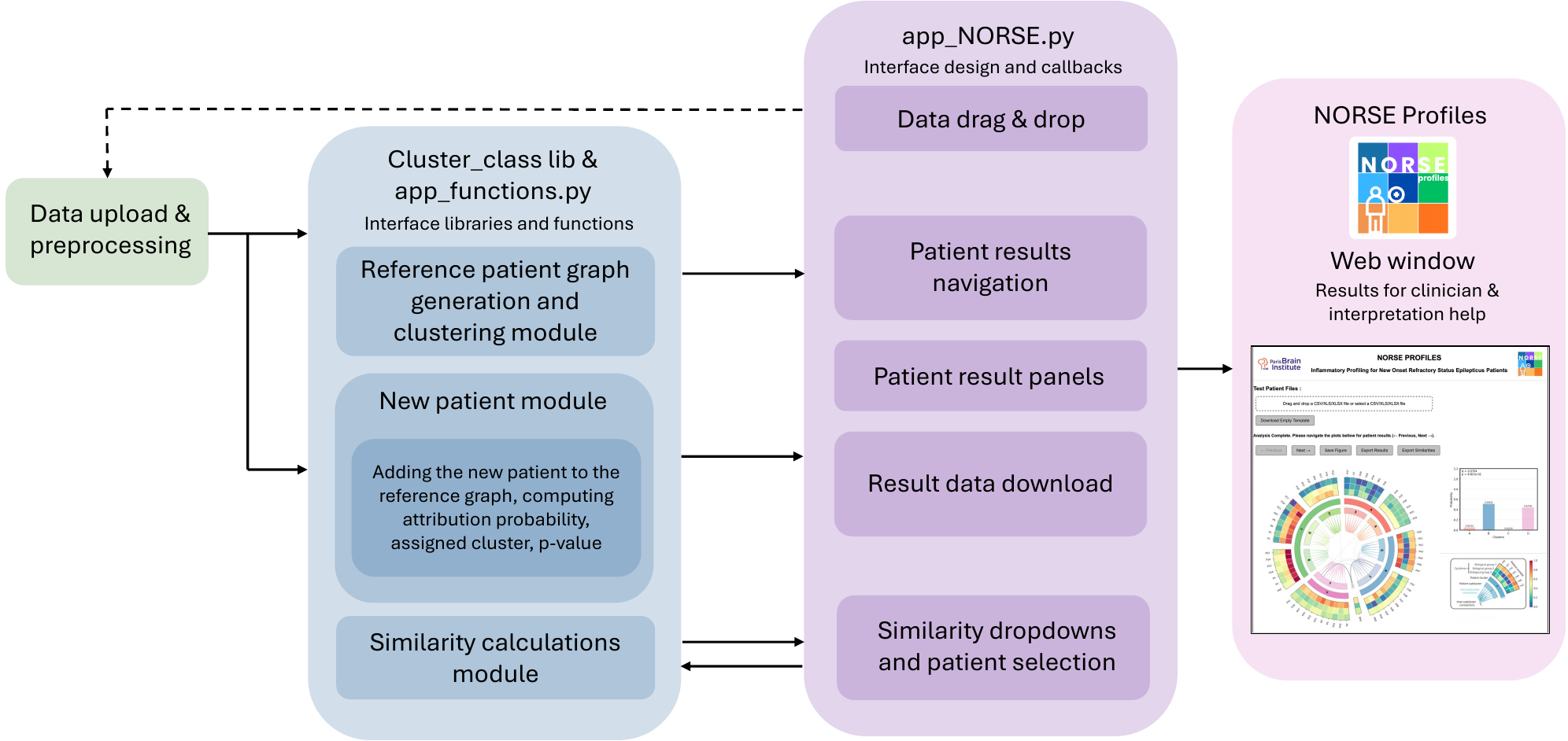}
\caption{The interface with the proposed pipeline. From left to right: data upload and preprocessing (green), reference patient graph and clustering modules, new patient module, optional similarity calculations module (blue), interface design and callbacks for navigation and control of patient results (purple), web window (pink).}
\label{figure3}
\end{figure*}

Ultimately, a user-friendly drag-and-drop interface (illustrated in Fig.~\ref{figure3}) was developed to enable the upload of spreadsheet files containing 96-cytokine panel measurements and to launch the analytical workflow described above. For each uploaded patient, the application projected the corresponding inflammatory profile onto the reference graph, estimated community and cluster membership probabilities, identified the most probable cluster assignment, computed the associated attribution p-value, and generated graphical representations of the patient’s position relative to the reference cohort. The platform additionally provided the described methods for quantifying biological similarity between pairs of samples. Results were generated automatically and displayed through a set of concise visualizations designed to facilitate interpretation. Multiple patients could be analyzed simultaneously from a single input file. To support clinical decision-making, the interface incorporated dedicated information panels summarizing the biological signatures, clinical outcomes, and therapeutic considerations associated with each inflammatory cluster. NORSE Profiles was implemented in Python 3.9 using the Dash framework (version 3.2.0) for web-based application development.

\section{Results}
\subsection{Attribution probabilities for new patients}
\begin{figure}[h]
\centering
\includegraphics[width=1.00\textwidth]{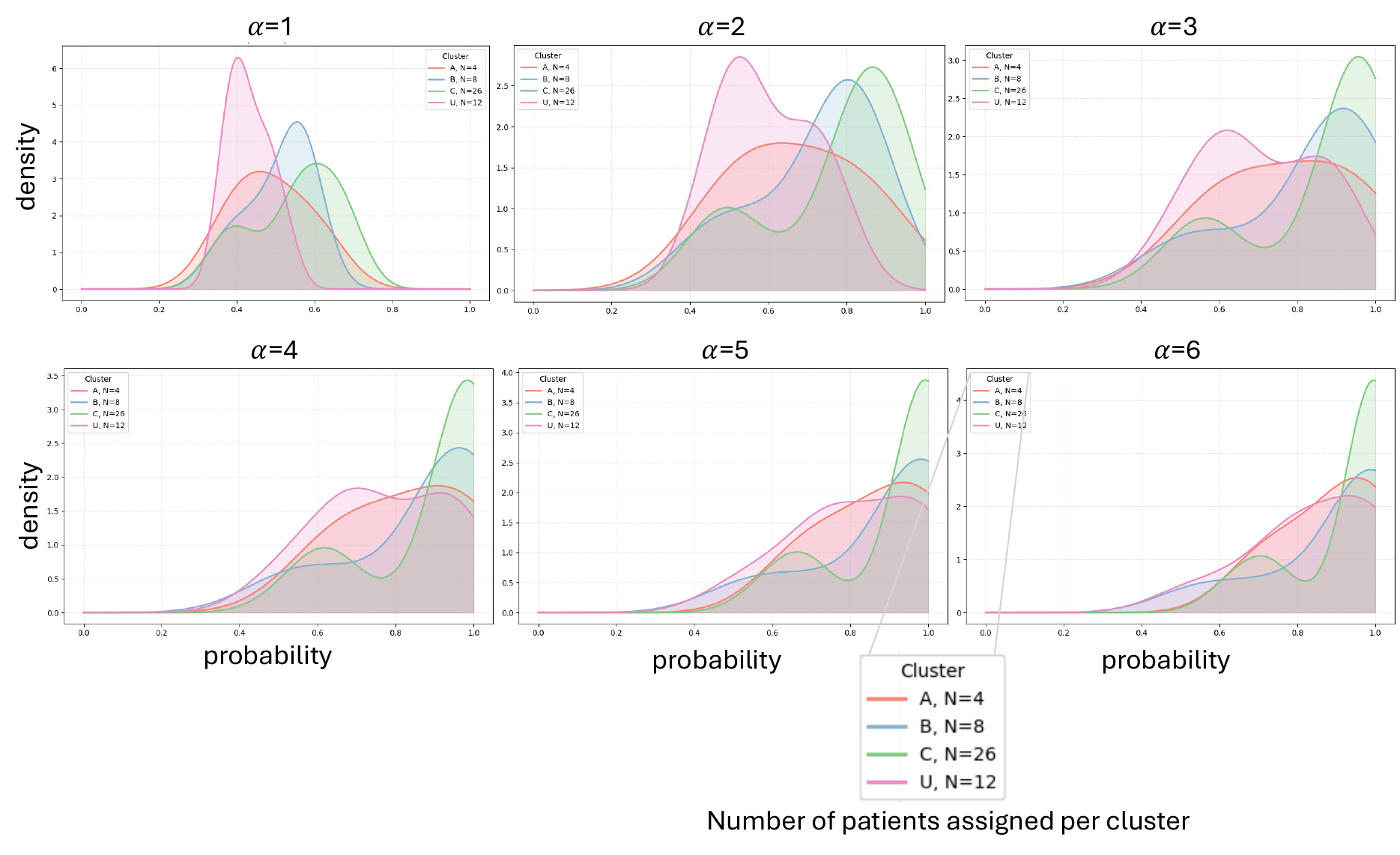}
\caption{Evolution of the attribution probability distributions per cluster in the SP model. The curves depend on the number of new NORSE patients assigned per cluster (N), indicated in the box bellow the plots.}
\label{figure4}
\end{figure}

The implementation of the SP model allowed the quantification of new patient assignments to our reference clusters A, B, C and U. The choice of the SP model and the parameter value $\alpha$ was set to improve clinical interpretability of attribution results, without introducing bias. Fig.~\ref{figure4} illustrates the evolution of attribution probability distributions per cluster in the SP model. We observed that larger values of $\alpha$ led to more confident cluster assignments (probabilities closer to one). For $\alpha=4$, attribution probabilities were confidently distributed although not excessively binarized, yielding improved interpretability for clinical application. Given that our reference graph $G$ was entirely connected, the use of the SP model to estimate attribution probabilities to every cluster was always possible. 

\subsubsection{Attribution probability distributions}
The three attribution models developed for new-patient classification, the shortest-path (SP), closest-neighbours (CN), and Euclidean-distance (EU) approaches (Supplementary Methods), yielded highly consistent results in the NORSENEW validation cohort. Overall, 48 of 50 patients (96\%) were assigned to the same inflammatory cluster by all three methods, whereas complete agreement was observed between the SP and EU models (50/50 patients, 100\%). In light of this strong concordance, the SP model was retained for all subsequent analyses and integrated into the \textit{NORSE Profiles} platform. In addition to its excellent classification agreement, the SP model offers the advantage of explicitly incorporating graph-topological information and allows continuous tuning of the balance between local and global network relationships through the parameter $\alpha$.

The SP model offered the additional advantage of incorporating a broader representation of the graph topology. In contrast, the CN model relied exclusively on local connectivity patterns, considering only the four nearest neighbours of the newly added patient. This methodological difference likely accounts for the two discordant cluster assignments observed across attribution models. Notably, the assignment probabilities associated with these discrepant cases were substantially lower than the average assignment probability observed for the remaining patients. Specifically, the CN-derived probabilities of assignment to Cluster B were 0.414 for patient P87 and 0.721 for patient L1, compared with a mean assignment probability of 0.870 across all other patients. These findings suggest that the discordant classifications occurred in cases with intrinsically lower assignment confidence and therefore do not materially affect the overall concordance between attribution methods.  

\subsubsection{Precision of new patient attributions}
\begin{figure}
\centering
\includegraphics[width=0.80\textwidth]{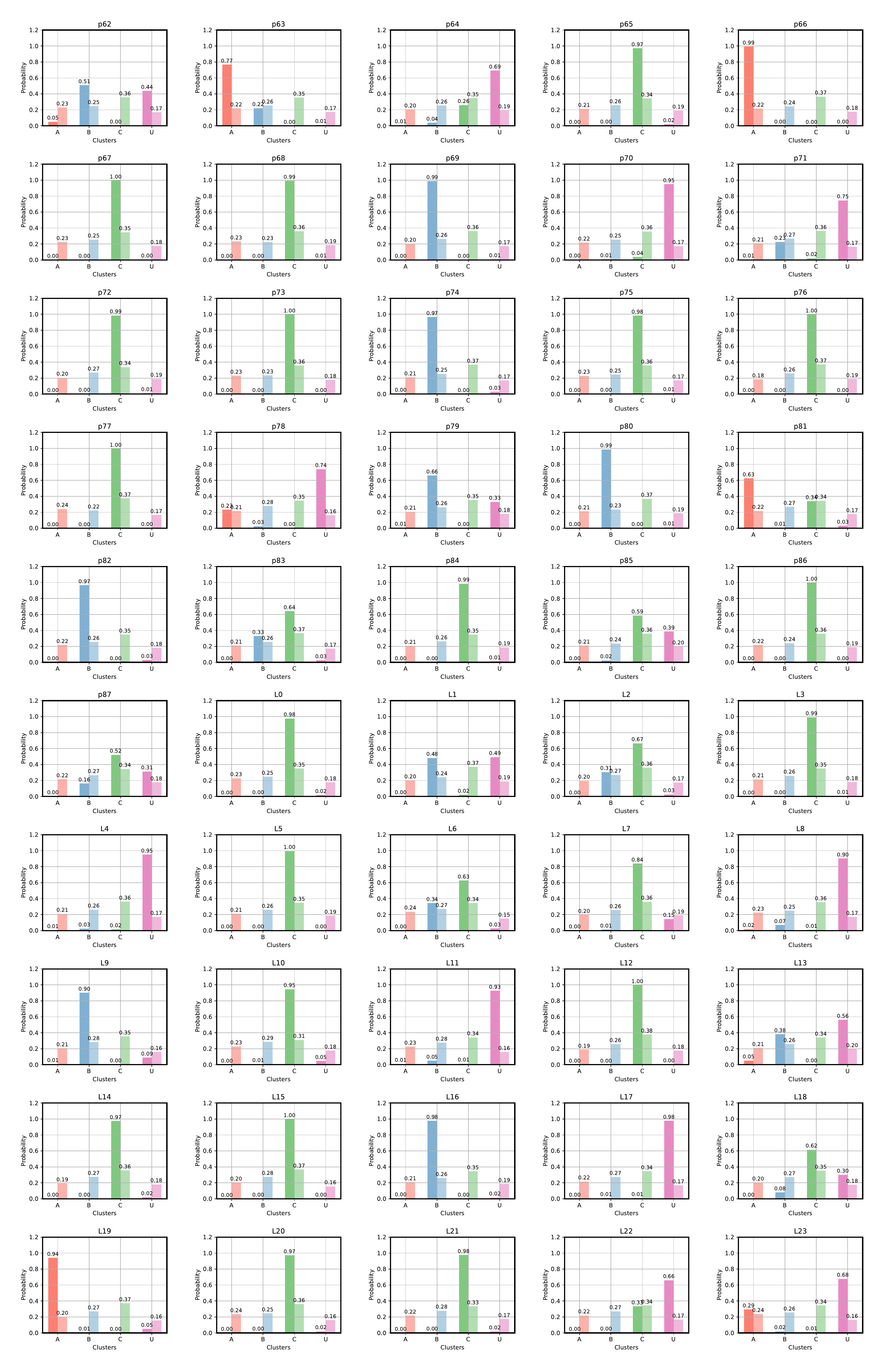}
\caption{ Attribution probability distributions in the SP model per cluster, per patient of the NORSENEW cohort ($\alpha=4$). Observed probability for each patient is represented by the opaque bar (left), and random probabilities are represented by transparent bars (right).}
\label{figure5}
\end{figure}

To statistically characterize the precision of our new patient assignments to clusters with the NORSENEW dataset, we tested the null hypothesis that newly added patients would not be assigned to clusters A, B, C or U with greater precision than would be expected if they were added randomly to the reference graph $G$. As described in section 2.4.1, we initially considered that the null distributions of cluster-assignment probabilities might depend on the specific patient being randomized during the Monte Carlo procedure. However, analysis of the simulated data demonstrated that the distributions of assignment probabilities under the null hypothesis were largely independent of the identity of the newly added patient (Fig.~\ref{figure5}). This observation justified the use of pooled estimates of random assignment probabilities for each cluster.  

Consistent with expectations, the average null assignment probabilities closely matched the relative frequencies of the corresponding clusters in the reference cohort (N=62). Specifically, the estimated null probabilities were proportional to the cluster prevalences: Cluster A, $13/62=0.21$; Cluster B, $16/62=0.26$; Cluster C, $22/62=0.35$; and Cluster U, $11/62=0.18$. The concordance between the empirically estimated null probabilities and the observed cluster frequencies in the reference cohort provides independent support for the validity of the Monte Carlo procedure used to characterize the null distribution of cluster assignments.  

\begin{figure*}
\centering
\includegraphics[width=0.80\textwidth]{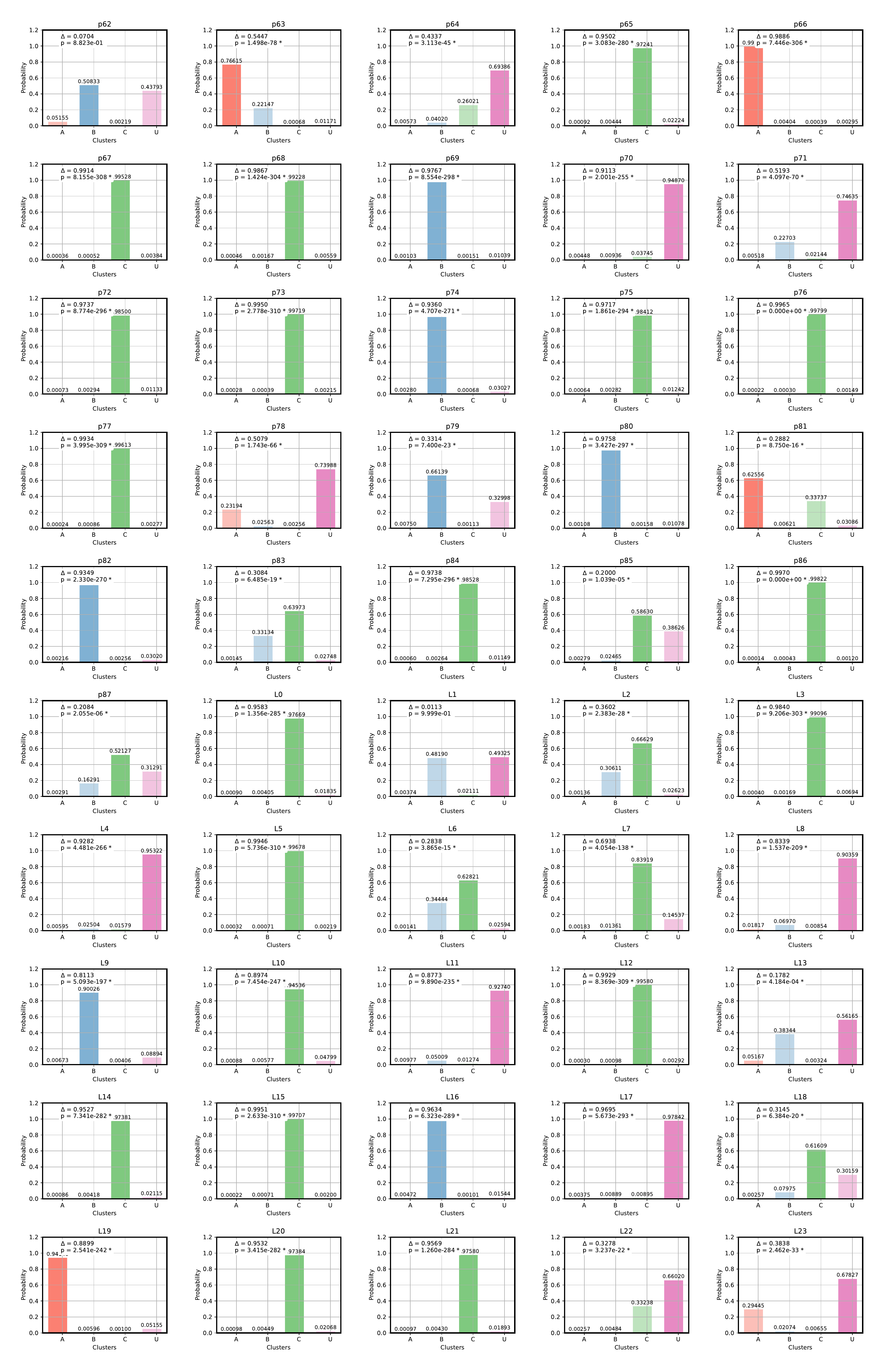}
\caption{Attribution probabilities in the SP model per cluster, per patient of the NORSENEW cohort ($\alpha=4$). Cluster assignment can be derived from the highest bars for each patient. Estimated $\Delta$ measures as well as p-values associated to the $\Delta$-test results figure in each sub-plot (the * indicates significance of attribution for $p<0.05$).}
\label{figure6}
\end{figure*}

Results of the $\Delta$-test are presented in Fig.~\ref{figure6} and yielded one significance value for each patient of the NORSENEW dataset. Using this approach, 48 of 50 patients (96\%) could be assigned to a cluster with statistical significance. The two non-significant cases corresponded to patients P62 and L1 (see Fig.~\ref{figure6}), both of whom exhibited highly similar assignment probabilities across two competing clusters.

Patient p62 was assigned to cluster B with an assignment probability of $p_{{SP}_B} =0.50833$, while the probability of assignment to Cluster U was only marginally lower $(p_{{SP}_U}=0.43793)$. Similarly, patient L1 was assigned to cluster B with a probability of $p_{{SP}_B}=0.48190$ whereas the probability of assignment to cluster B was nearly identical ($p_{{SP}_U}=0.49325$). In both cases, the minimal difference between the two highest assignment probabilities resulted in a non-significant $\Delta$-Test, indicating substantial uncertainty regarding cluster membership. These findings suggest that patients P62 and L1 occupy boundary regions of the graph, where inflammatory profiles exhibit features shared by multiple clusters. 

We further investigated the influence of the parameter $\alpha$ on the performance of the $\Delta$-test and the resulting cluster assignments. We observed that, for values of $\alpha$ within the range $\left[1,4\right]$, the estimated p-values varied substantially, reflecting the sensitivity of the attribution procedure to the relative weighting of local versus global graph topology. The highest number of statistically significant assignments (48/50 patients) was achieved for $\alpha \geq 4$. Overall, increasing $\alpha$ progressively enhanced the influence of nodes in close topological proximity to the newly added patient, while reducing the contribution of more distant regions of the graph. Consequently, larger values of $\alpha$ yielded increasingly stringent cluster assignments. For $\alpha \geq 4$, assignment probabilities became progressively polarized, approaching either unity or zero (Supplementary Methods). This effect was observed for all patients, with the exception of P62 and L1, whose assignment probabilities remained intermediate because of their proximity to cluster boundaries. The choice of $\alpha=4$ therefore represented an optimal compromise, preserving sufficient granularity in the estimated p-values while maximizing both the statistical significance and confidence of new-patient cluster assignments.

\subsection{New patient profiling for clinical interpretability}
Our method was developed as an intuitive front-end clinical decision-support tool for neurologists and intensive care physicians. It aims to facilitate the interpretation of inflammatory profiles in NORSE from high-dimensional cytokine data, ultimately enabling patient assignment to distinct clusters. We evaluated the method using patient data from the NORSENEW, RSE, CHRONIC, ENCEPHALITIS, NORSELONGI, and HCONTROL datasets. 

\begin{figure}[htbp]
    \centering
    \includegraphics[width=0.65\columnwidth]{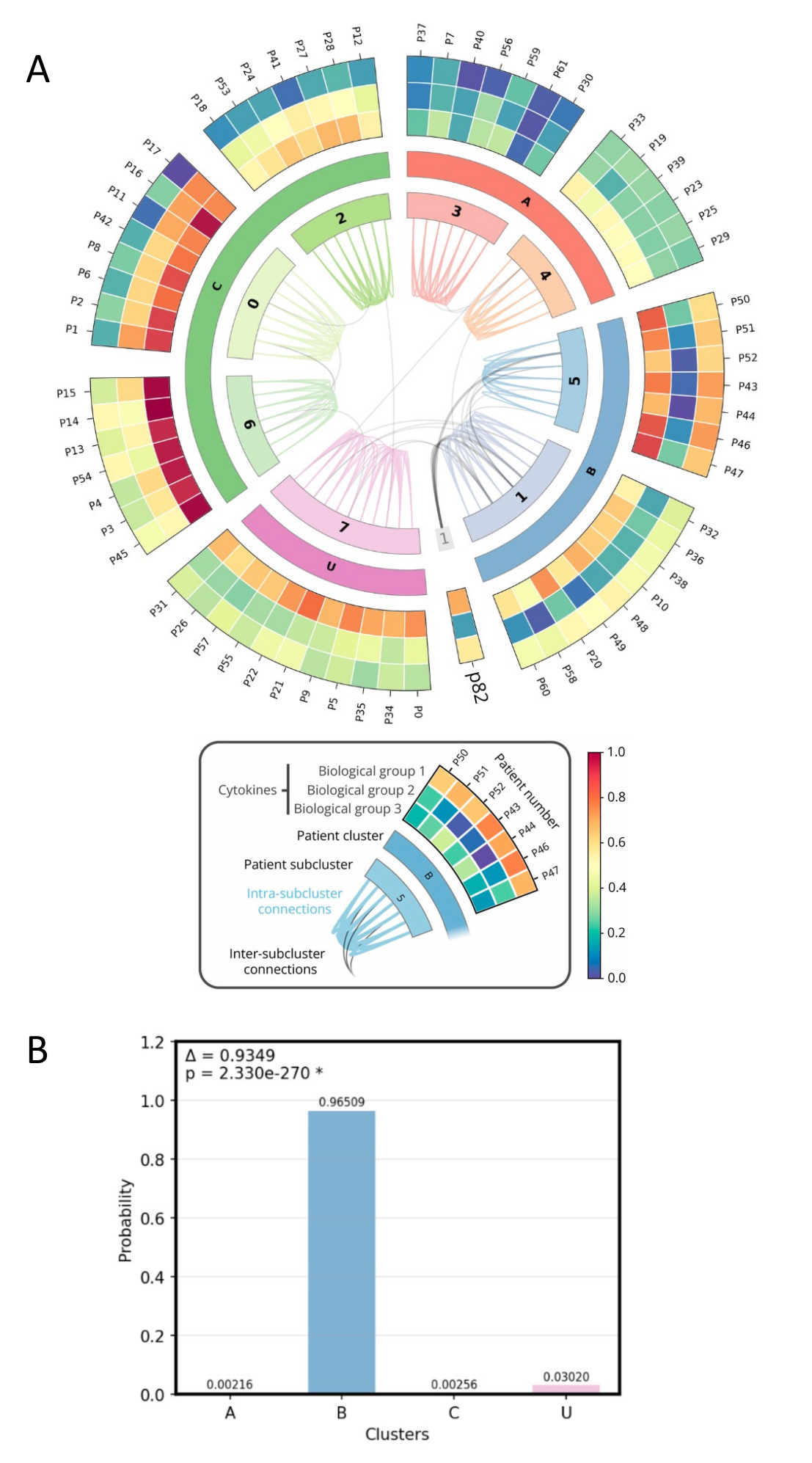}
    \caption{Inflammatory profiling results for patient P82 assigned to cluster B with an attribution probability $p_{{SP}_B}=0.96509$, and $p=2.330\times10^{-270}$. \textbf{A.} Circular plot positioning P82 among the reference inflammatory profiles of NORSEREF as originally estimated in \cite{guillemaud_identification_2025}. \textbf{B.} Attribution probability distribution and p-value panel.}
    \label{figure7}
\end{figure}

Fig.~\ref{figure7} presents our interface output results for an illustrative case, patient: P82. The analysis indicates that patient P82 is assigned to cluster B with a probability $p_{{SP}_B}=0.96509$ and a very low p-value ($p=2.330\times10^{-270})$. Assignment to cluster B is indicative of an acute inflammatory profile associated with a poor prognosis. Consistent with these results, clinical records indicate that patient P82 presented with a severe form of NORSE, was refractory to administered treatments, and ultimately died during hospitalization. Assignment to cluster B suggests that the patient has an acute inflammatory profile, with worst prognostics. Consistently with these results, clinical observations for patient p82 indicated that the patient suffered from a severe NORSE condition, unresponsive to the treatment administered; the patient passed away during his hospitalisation. 
Furthermore, across all patients in the NORSENEW dataset, we observed no discrepancies between the output of our results and the available clinical observations. 

\subsection{Cluster-tracking inflammatory profiles}
We here summarize the results of our longitudinal study conducted to estimate the similarity metric $s_{EU}$ across a cohort of NORSE patients who had been sampled and analysed at two different time points $t_1$ and $t_2$. Table~\ref{table2} reports the cluster assignment of samples at time points $t_1$ and $t_2$, the corresponding attribution probability, and the significance level according to our $\Delta$-Test. 

Overall, 14 our of 24 NORSE patients exhibited a change in cluster assignment between the two time points. Among these patients, no clear correlation was observed between the similarity metric $s_{EU}$ and the time interval separating the two samplings ($t_2-t_1$). The largest interval ($t_2-t_1$= 24 days) was observed for patient L9, who remained consistently assigned to cluster B at both time points. Conversely, some patients exhibited cluster transitions after only a one-day interval.

\begin{figure}[h]
    \centering
    \includegraphics[width=0.6\columnwidth]{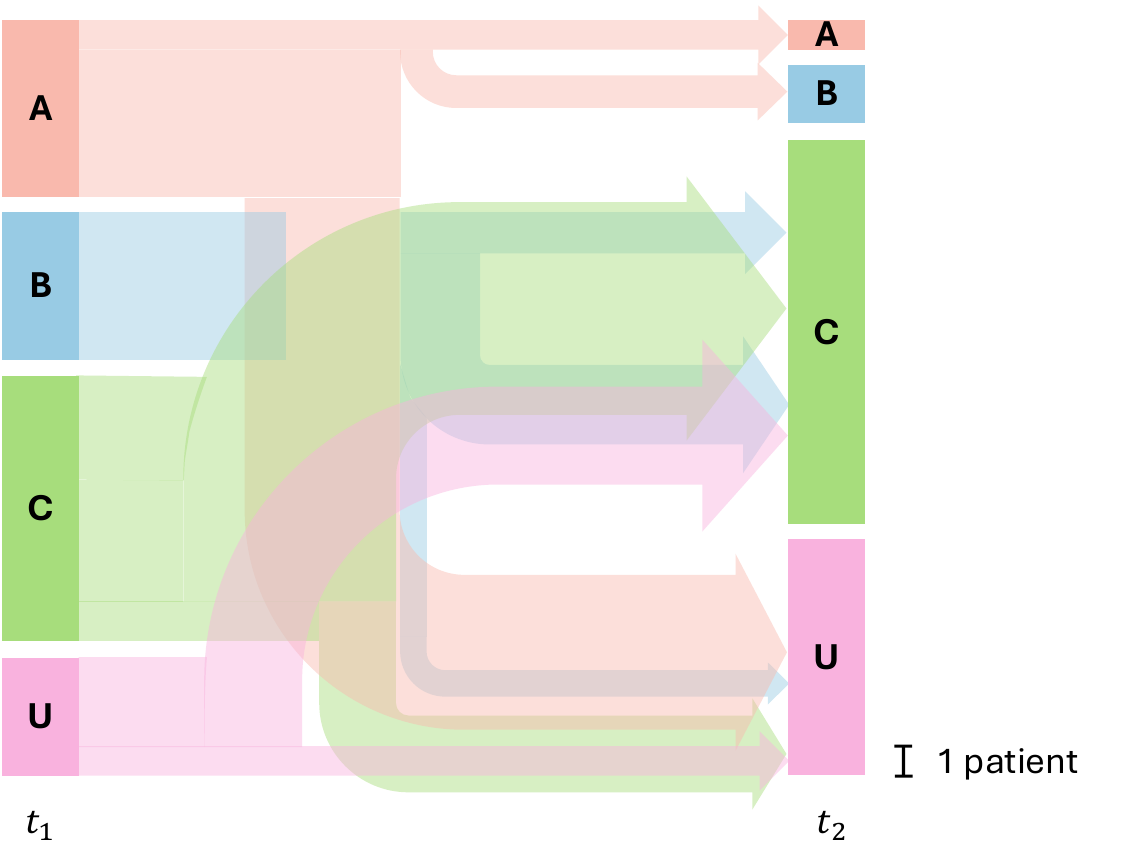}
    \caption{Summary of cluster transitions (arrows) observed in the longitudinal study on NORSELONGI data. Clusters are represented by the vertical bars at times $t_1$ and $t_2$. The scale for one patient is provided in legend next to the figure. The size of the arrows are proportional to the number of transitioned patients.}
    \label{figure8}
\end{figure}

Notably, patients who remained stable in their cluster assignments were associated with substantially lower p-values in our model (typically $<4.054 \times 10^{-138}$). The most frequent transitions were from cluster A to cluster U (4/14 cases). Patients from cluster C remained most stable in their cluster assignments (7/10), as illustrated in Fig.~\ref{figure8}.

It is important to note that the time points used in this study were not associated with a standardized sampling procedure. This may have affected the results, as they also depend on patient inclusion dates and patient-specific treatment trajectories (see Discussion). We investigated potential biological batch effects by analyzing inflammatory profiles of other patients from the same biological batch, because of the stronger homogeneity of profiles sampled at $t_2$ as compared to $t_1$ (see Table~\ref{table2} and Supplementary Methods).

\begin{table*}[t]
\centering
\includegraphics[width=1.00\textwidth]{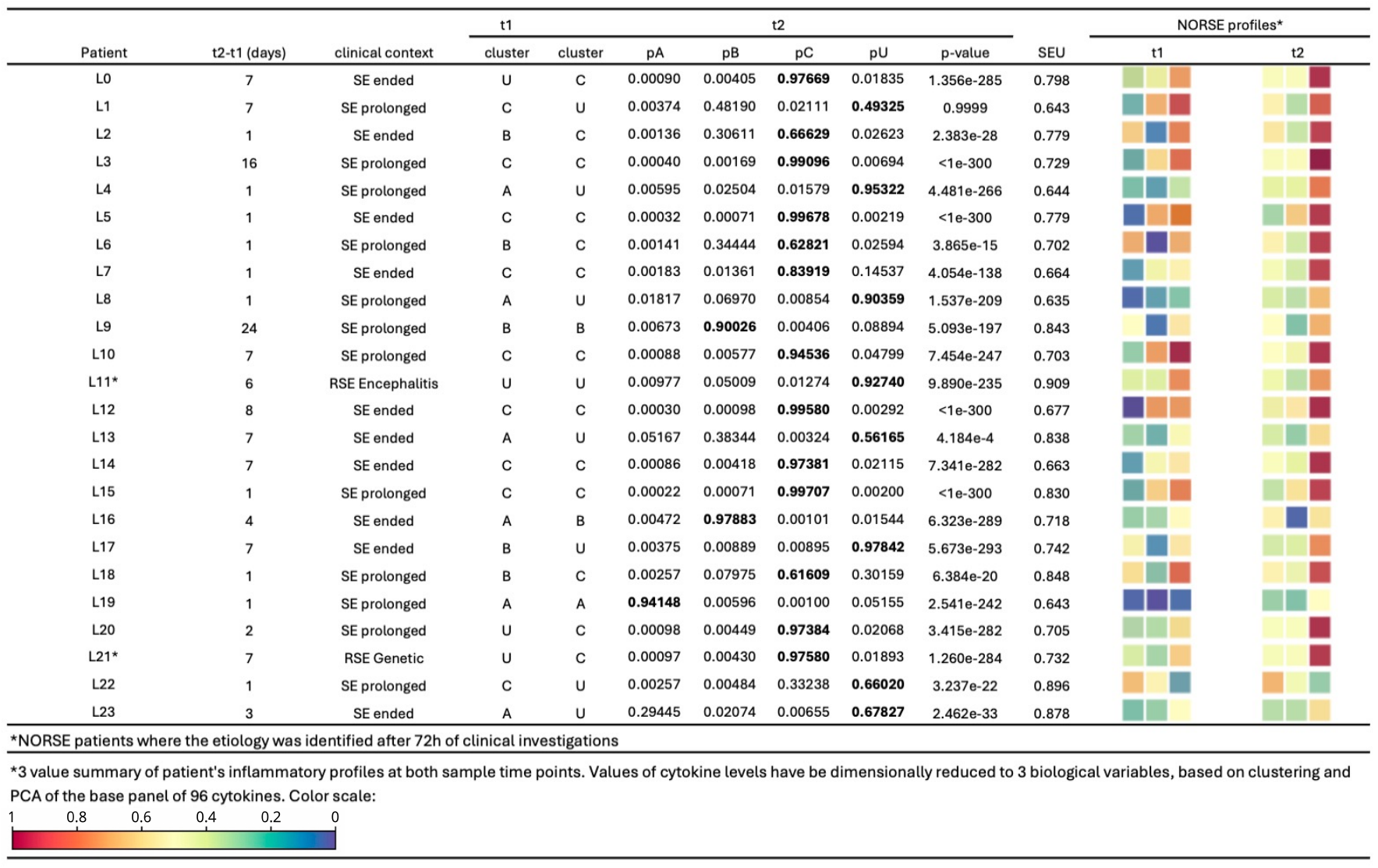}
\caption{Summary of the longitudinal study on the 24 patient NORSELONGI dataset. Patients were sampled at 2 time points $t_1$ and $t_2$, eventually separated by a different number of days. Bold attribution probabilities refer to the assigned cluster probability.}
\label{table2}
\end{table*}

\subsection{Influence of reference cohort variations}
\begin{figure*}
    \centering
    \includegraphics[width=1.00\textwidth]{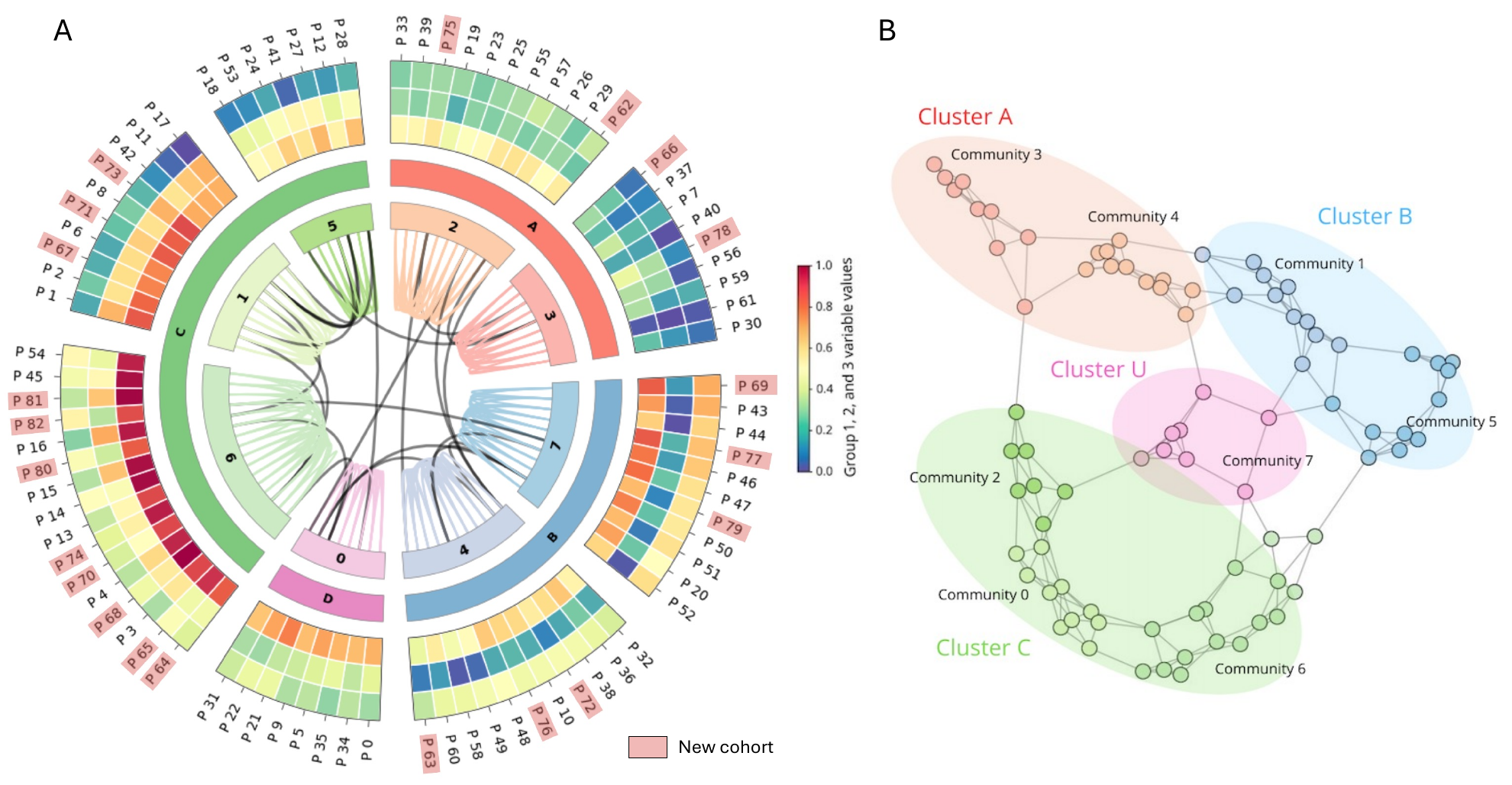}
    \caption{Extended reference cohort study.\textbf{ A.} Cluster assignment with the extended cohort of 82 NORSE patients. \textbf{B.} Clustered graph structure with the extended cohort.}
    \label{figure9}
\end{figure*}

Assuming that the reference graph $G$ and its partition $C$ into clusters A, B, C and U are fixed and biologically valid \cite{guillemaud_identification_2025}, we assessed the robustness of our results in the context of prospective extensions of the NORSE reference cohort. While preserving the biological variable groups 1, 2 and 3, we repeated the clustering procedure on an extended reference cohort comprising both NORSEREF and NORSENEW patients with complete 96-cytokine panels (N=82 NORSE patients, compared to N=62 in the original cohort). 

As shown in Fig.~\ref{figure9}, the graph-based clustering implementation is highly stable with respect to cohort extension. We recovered the same eight communities (i.e. partition $C$) and the same higher-level clusters A, B, C and U, consistent with the original community composition: A $= \{C_3, C_4\}, \ \text{B} = \{C_1, C_5\}, \ \text{C} = \{C_0, C_2, C_6\}, \text{and} \ \text{U} = \{C_7\}$. Overall, the graph structure illustrated in Fig.~\ref{figure9}-B remained consistent with the reference graph, in which the undetermined cluster U acts as a hub connecting clusters A, B, and C. 

In total, 95\% of patients retained their original cluster assignment, and the distribution of NORSE patients across clusters remained comparable to that observed in the reference graph-based clustering ($N_{A_{ref}}=21\%,N_{A_{82}}=24\%,N_{B_{ref}}=26\%,N_{B_{82}}=27\%,N_{C_{ref}}=35\%,N_{C_{82}}=39\%$). Notably, the only three patients (P26, P55, P57) whose assignments changed were reassigned from the undetermined cluster U to cluster A in the extended cohort analysis. This observation suggests that extending the reference cohort may improve cluster resolution and assignment confidence within our framework (see Discussion).

\section{Discussion}

In this study, we demonstrated that graph-based statistical models can be effectively implemented to characterize the inflammatory profiles of newly diagnosed NORSE patients. By using patient graph-based clustering approaches, we further advanced efforts to elucidate the biological heterogeneity underlying NORSE through the use of accessible and minimally invasive biomarkers. The proposed framework enables the integration and interpretation of high-dimensional cytokine data within a clinically meaningful structure, thereby facilitating the identification of distinct inflammatory phenotypes. 

Moreover, the methodology introduced in this work allowed for robust patient assignment and quantitative assessment of attribution confidence, substantially improving the clinical interpretability of complex inflammatory profiles. By providing an objective characterization of immune dysregulation patterns, our approach has the potential to support clinical decision-making in NORSE and to inform the selection and adaptation of immunomodulatory therapies according to the specific inflammatory state of individual patients. 

Our study found relative stability through cohort extension in the reference graph $G$ structure and patient-cluster composition obtained through the computationally efficient Louvain graph-based algorithm \cite{blondel_fast_2008}. Other graph-based clustering approaches however exist \cite{girvan_community_2002, newman_finding_2004, pons_computing_2006, newman_finding_2006}, including very recent potentially competitive alternatives \cite{meena_graph_2025}. A limit of our approach is its reliance on the biological validity and stability of the initially defined clusters A, B, C and U. Nevertheless, the robustness analyses conducted in this study successfully reproduced the original clustering structure and patient assignments, thereby supporting the reliability of the proposed framework and corroborating previous findings \cite{guillemaud_identification_2025}. Furthermore, we were able to prematurely verify the biological validity of the patient-clustering model through the concomitance of clinical observations we had at disposal, and cluster assignment to characteristic inflammatory profiles. 

Future prospective clinical studies evaluating tools such as \textit{NORSE Profiles} will be necessary to establish the clinical utility of this approach more rigorously, particularly with respect to treatment personalization and patient outcomes. Such investigations could determine whether cluster-guided therapeutic strategies improve functional recovery and disease managment in NORSE \cite{wang_precision_2023}.

The implementation and power of our statistical $\Delta$-Test was partially limited by the amount of data we had at disposal when dealing with rare conditions such as NORSE. Approximations of our $\Delta$ statistic distribution to a normal distribution enabling parametric testing may have to be replaced by more empirical approaches in the future. In this case, the proposed methodology can easily be adapted with a cumulative distribution function, and a non-parametric test to estimate p-values instead. Moreover, our model parameter value choices were made in large considerations with clinical interpretability of results. 

The approach’s potential for tracking inflammatory profiles showed inconclusive results. We did observe patient transitions from their initially assigned cluster to another from $t_1$ to $t_2$, suggesting that cluster-tracking models borrowed from social-network fields might be relevant to characterizing dynamic patient profiling and follow-up over time \cite{lambert_tracking_2023, greene_tracking_2010}. However, these transitions were difficult to interpret due to the lack of a standardized collection of sample procedure. Time points $t_1$ and $t_2$ weren’t unified across patients, and some individuals would have likely received immunomodulatory treatments in between two samplings, introducing further bias for direct interpretation of inflammatory profile evolution. 

By standardising the sampling protocol in future studies, the longitudinal monitoring of inflammatory profiles could be strengthened and supported by our similarity measure. This measure we incorporated to the study is also Euclidean distance-based, but alternatives through graph-based calculations (SP) or different distance metrics may be tested \cite{irani_clustering_2016}. Other limitations related to the serum sample collection processes may have introduced bias into the profile results, which could then depend on minor biological batch effects. This study therefore assumes proper initial sampling and handling of patient samples, in accordance with literature recommendations \cite{hanin2023review}. 

Currently, our framework was designed to operate on an extended panel of 96 cytokines, which is not always rapidly available in a clinical setting. Current imputation strategies have been integrated to moderate the effect of missing data; however, to guarantee future accessibility, the methodology would require adaptation to a reduced panel of cytokines. Additionally, our framework has been shaped around the use of more intra-correlated cytokines groups 1, 2 and 3 based on the patient datasets we had at disposal. The composition of our biological variables may be subject to slight evolutions in the future, if further data is generated. Indeed, we observed some slight variations in cytokine group composition of our biological variables with the use of our extended cohort. Efforts should nonetheless focus on maintaining biologically interpretable cytokine variables for clinical applications, as previous works have also encouraged the identification of cytokine clusters independently from patient clusters \cite{polley_identification_2023}. 

Considering past findings of differentially elevated cytokines and chemokines in the cerebrospinal fluid (CSF) of paediatric FIRES patients, the model we here implemented from serum samples could be extended to CSF samples \cite{https://doi.org/10.1111/epi.16275}. As compared to noninflammatory neurological disorders and encephalitis cases, FIRES patient CSF showed elevations of cytokines already included in the panel we were working with (TNF-$\alpha$, CXCL9, IP-10, I-TAC, IL-6, MCP-1, MIP-3$\beta$ and GRO-$\alpha$) \cite{https://doi.org/10.1111/epi.16275}. Moreover, our methodology may have potential applications to a wider range of diseases which can also be characterized through cytokine-level based inflammatory profiling, including for studying personalized responses to immunotherapies \cite{xu_circulating_2025}.

\section{Conclussion}
This study introduced a novel graph-based statistical tool for inflammatory profiling in NORSE. We implemented this approach to personalized immunomodulatory strategies for patients with cNORSE, where few information is available at the time of the SE to guide the clinician and the risk of acute neurological sequelae remains significant. As more samples from affected patients are collected and analyzed, it seems possible to confirm the robustness of the initially developed approaches and to adapt the methods to future clinical, and locally accessible, practice. 

\section*{Ethics}
This study follows ethical and academic integrity guidelines. The study was approved by the Paris Pitié-Salpêtrière Hospital (APHP, COLETTE and Inserm, TIPI) and Yale University (NORSE/FIRES biorepository, IRB \#1511016840 and \#2000031611). Informed consent was obtained from all patients or legally authorized respresentatives following the Declaration of Helsinki.

\section*{Authors’ Contributions}
L.D.: investigation, methodology, results visualization, writing—original draft, and editing; 
M.G.: data curation, investigation, methodology, writing—original draft;
V.N.: data curation, writing—review and editing;
A.H.: data curation, methodology, writing original draft and editing;
M.C.: conceptualization, supervision, result visualization, writing—review and editing;
All authors gave final approval for publication and agreed to be held accountable for the work performed therein.

\section*{Funding}
This research received no external funding.

\section*{Conflict of interests}
The authors declare no conflict of interest


\bibliographystyle{elsarticle-num} 
\bibliography{references}

\newpage
\appendix
\renewcommand{\thefigure}{S\arabic{figure}}
\renewcommand{\thetable}{S\arabic{table}}
\setcounter{figure}{0}
\setcounter{table}{0}
\setcounter{section}{0}

\section{Supplementary Methods}
\subsection*{The CN model}
To attribute clusters to a newly added patient to graph 
\[
\widetilde{G}=G\cup\left(V_{N+1},\ E_{N+1},\ W_{N+1}\right), 
\]
we developed a second statistical approach where patient $N+1$ is assigned to the most frequently represented cluster in his neighbourhood formed by his $x$ closest neighbours (CN). This CN model was compared and selected amongst two other tailored attribution models (SP and EU). Using the same notations introduced in the main text, we additionally define $\Pi\left(j\right)=k$ the assignment of node $j$ to community $C_k$, for $k\in\left\{1,\ 2,..,\ K\right\}$. For each community $k$, the probability of new patient N+1 being attributed to that community is described as:
\begin{equation}
P_{N+1,k}=\left\{\begin{matrix}\frac{\sum_{j\in V_{N+1},\mathrm{\Pi}\left(j\right)=k} w_{N+1,j}}{\sum_{j\in V_{N+1}} w_{N+1,j}},&\mathrm{if\ }C_k\in\mathrm{\Pi}\left(V_{N+1}\right),\\0,&\mathrm{else}\end{matrix}\right.
\tag{S1}
\end{equation}
where the assigned community can be derived as $\max_{k\in \left\{1,\ldots,K\right\}}{P_{N+1,k}}$, and $V_{N+1}$ represents the neighbourhood of patient $N+1$. The normalization term yields a probabilistic measure of community membership for the newly added node.

\subsection*{The EU model}
Similarly to SP and CN models, we developed a third statistical approach where a new patient $N+1$ is assigned to the spatially closest community from the clustered graph. In this case, we embedded the graph in a Euclidean space characterised by our three biological variables $(PC1_{group1}, PC1_{group2}, PC1_{group3})$. Let $d_{i,j}$ be the distance between the vectors corresponding to projected node $i$ and node $j$. We defined the attribution probability for node $N+1$ to community $C_k$, for $k\in\left\{1,\ 2,..,\ K\right\}$ in our Euclidean model (EU) such that: 

\begin{equation}
P_{N+1,k}=\frac{1}{K-1}\left(1-\frac{\sum_{j\in C_k} d_{N+1,j}}{\sum_{j=1}^{N}d_{N+1,j}}\right)   
\tag{S2}
\end{equation} 

\begin{figure*}[hbt]
\centering
\includegraphics[width=0.60\textwidth]{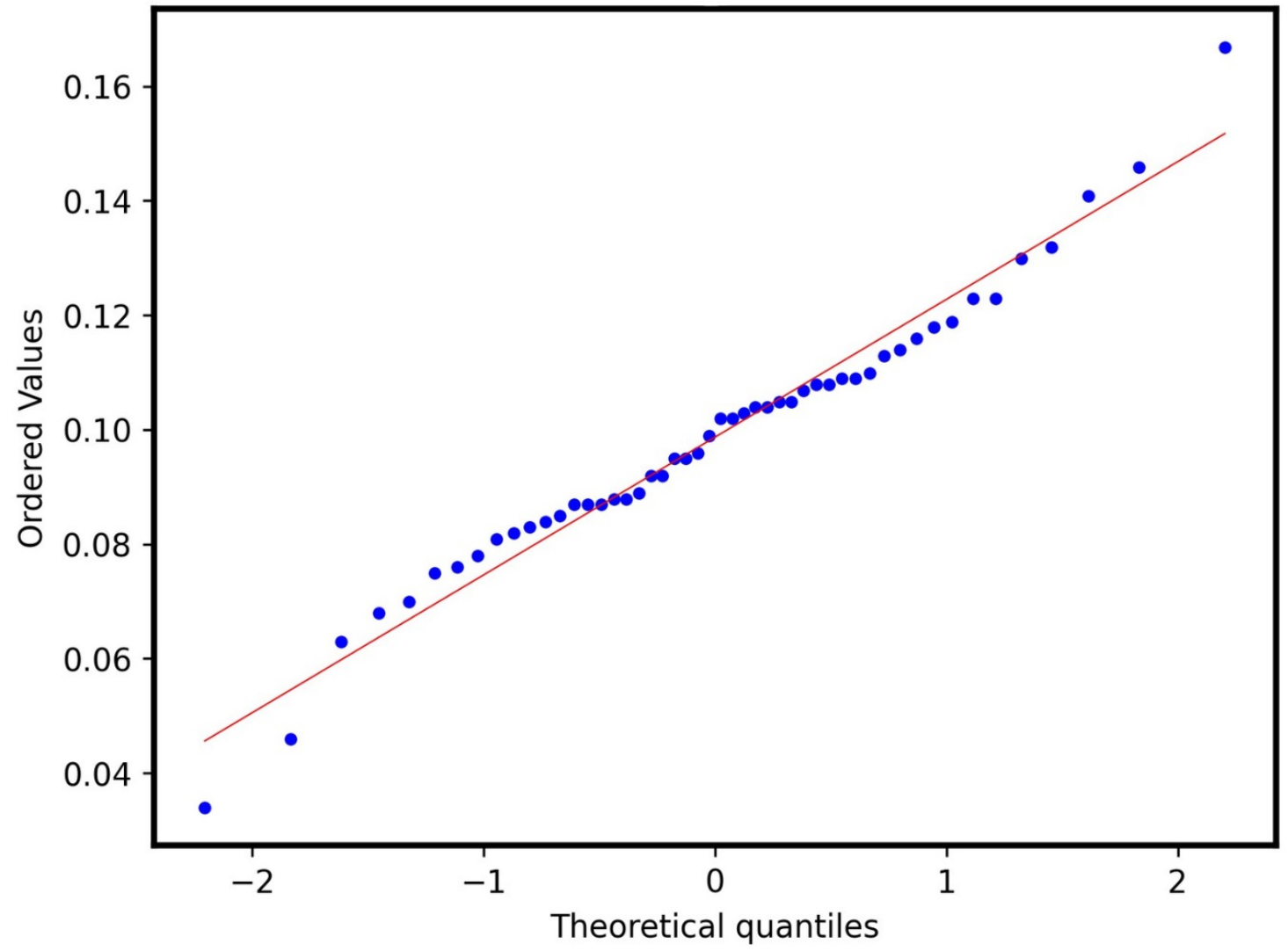}
\caption{The QQ distribution of the simulated $\Delta$.}
\label{figureS5}
\end{figure*}

\subsection*{Approximation of the $\hat\Delta$ distribution}
The distribution of $\hat\Delta$ in a random situation was derived from the Monte-Carlo simulations performed on patient data. The Shapiro-Wilk test was applied (W=0.980, $p=0.565$). We consequently did not reject the null hypothesis and further performed some verifications on the observed distribution’s tails (see Fig.~\ref{figureS5}). We observed with vigilance the behaviour of the tails depending on the number of iterations used in the simulation. We found a false positive rate of 0.06 for a significance bar at $p<0.05$ and 0.02 for $p<0$.01, with 1000 iterations. The sensitivity associated to the Monte-Carlo simulation noise was estimated at s=0.0069 for a bar at $p<0.05$ and at $s=0.0031$ for a bar at $p<0.01$.

\clearpage
\section*{Supplementary Figures and Tables}
\subsection*{Examples of patient results behaviour to $\alpha$ parameter variations in the $\Delta$-test}
\begin{figure}[h]
\centering
\includegraphics[width=1.00\textwidth]{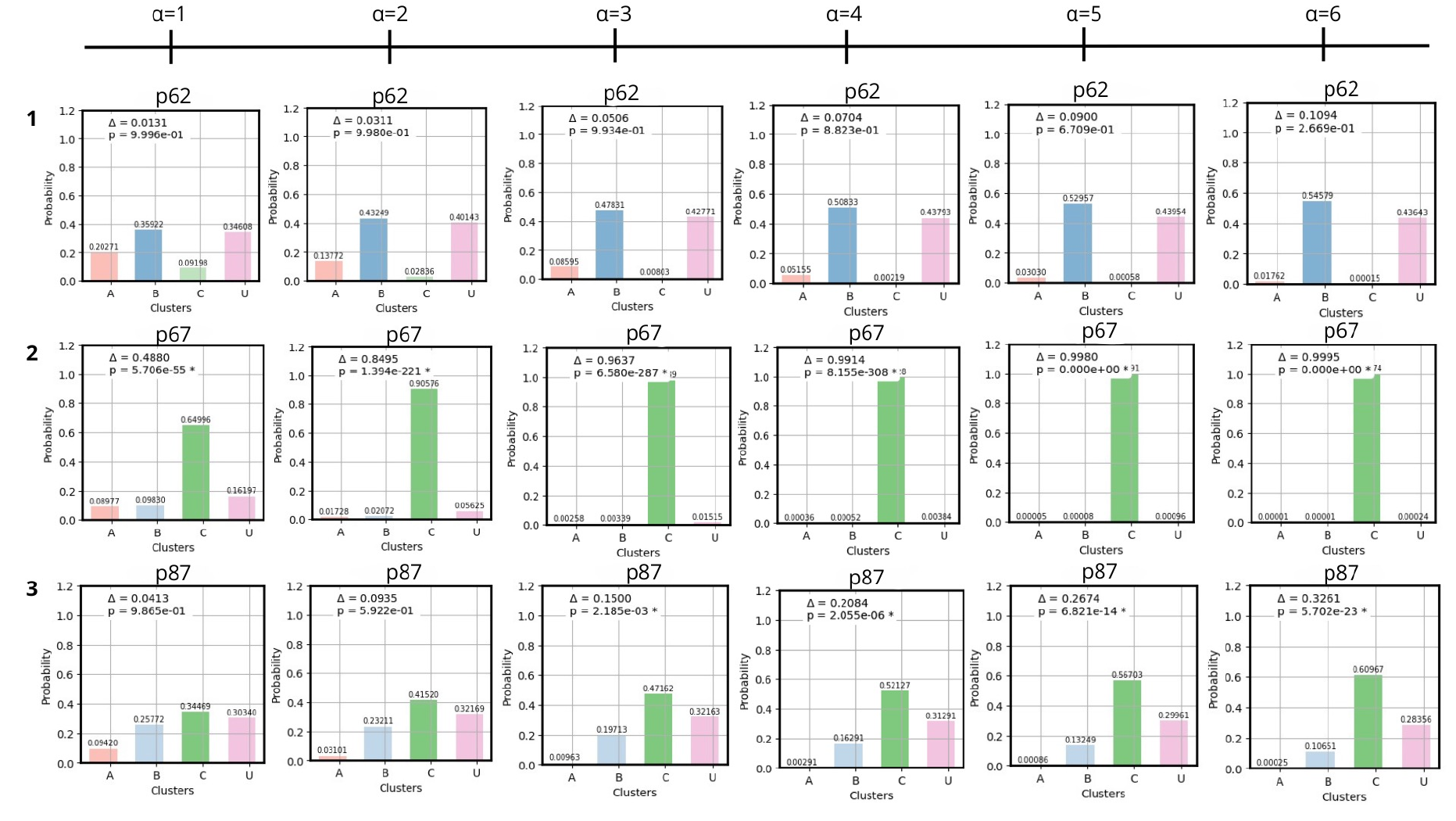}
\caption{Patient examples illustrating the effect of the parameter $\alpha$ in the SP model. \textbf{1.} Patient p62: non-significant, with decreasing p-values showing the progressive dominance of clusters B and U on clusters A and C, reducing nuance in probability estimations. \textbf{2.} Patient p67: highly significant even with $\alpha=1$, and progressive dominance of cluster C with the increase of parameter $\alpha$, also reducing nuance in probability estimations.\textbf{ 3.} Patient p87: only significant for $\alpha \geq 3$. }
\label{figureS6}
\end{figure}
\clearpage

\subsection*{The 96-cytokine panel}

\begin{table}[htbp]
\centering
\scriptsize

\begin{minipage}[t]{0.47\textwidth}
\centering
\begin{tabular}{|l|l|l|}
\hline
\textbf{Group 1} & \textbf{Group 2} & \textbf{Group 3} \\
\hline
IL-1RA          & MIP-1$\alpha$      & sFasL \\
GRO$\alpha$     & IL-3               & I-309 \\
GCP-2           & CXCL16             & IL-35 \\
MDC             & VEGF-A             & IL-23 \\
TARC            & IL-4               & IL-16 \\
ENA-78          & IL-7               & CCL28 \\
EGF             & GM-CSF             & 6CKine \\
PDGF-AB/BB      & Fractalkine        & IL-28A \\
sCD40L          & TNF$\beta$         & IFN$\omega$ \\
IL-10           & IL-13              & IFN$\beta$ \\
IL-15           & IFN$\gamma$        & IL-34 \\
M-CSF           & MCP-3              & LIF \\
sFas            & IL-22              & Eotaxin-3 \\
TPO             & IL-17E/IL-25       & IL-24 \\
MIP-1$\delta$   & IFN$\alpha$2       & IL-11 \\
SDF-1           & IL-9               & HMGB1 \\
IL-8            & IL-12p70           & Lymphotactin \\
MCP-1           & TSLP               & IL-21 \\
MIP-1$\beta$    & TGF$\alpha$        & IL-31 \\
APRIL           & IL-17F             & IL-29 \\
SCF             & IL-33             & \\
G-CSF           & IL-12p40          & \\
IL-6            & IL-5              & \\
\hline
\end{tabular}
\end{minipage}
\hfill
\begin{minipage}[t]{0.47\textwidth}
\centering
\begin{tabular}{|l|l|l|}
\hline
\textbf{Group 1} & \textbf{Group 2} & \textbf{Group 3} \\
\hline
IL-20           & sCD137            & \\
MIP-3$\alpha$   & Granzyme B        & \\
IP-10           & IL-1$\beta$       & \\
MIP-3$\beta$    & FGF-2             & \\
IL-18           & IL-2              & \\
BAFF            & IL-17A            & \\
FLT-3L          & IL-1$\alpha$      & \\
TNF$\alpha$     &                   & \\
MCP-4           &                   & \\
MIG/CXCL9       &                   & \\
MCP-2           &                   & \\
Granzyme A      &                   & \\
I-TAC           &                   & \\
Eotaxin         &                   & \\
IL-27           &                   & \\
BCA-1           &                   & \\
MPIF-1          &                   & \\
CTACK           &                   & \\
TRAIL           &                   & \\
Eotaxin-2       &                   & \\
RANTES          &                   & \\
Perforin        &                   & \\
PDGF-AA         &                   & \\
\hline
\end{tabular}
\end{minipage}
\caption{The 96-cytokine panel used in this analysis, sorted by biological groups of cytokines formed through correlation analysis (Eve Technologies’ Human Cytokine 96-Plex Discovery Assay (MilliporeSigma, Burlington, MA)).}
\label{tableS1}
\end{table}
\clearpage

\subsection*{Inflammatory profiling results for control of batch effects in the longitudinal study}
\begin{table}[h]
\centering
\includegraphics[width=0.80\textwidth]{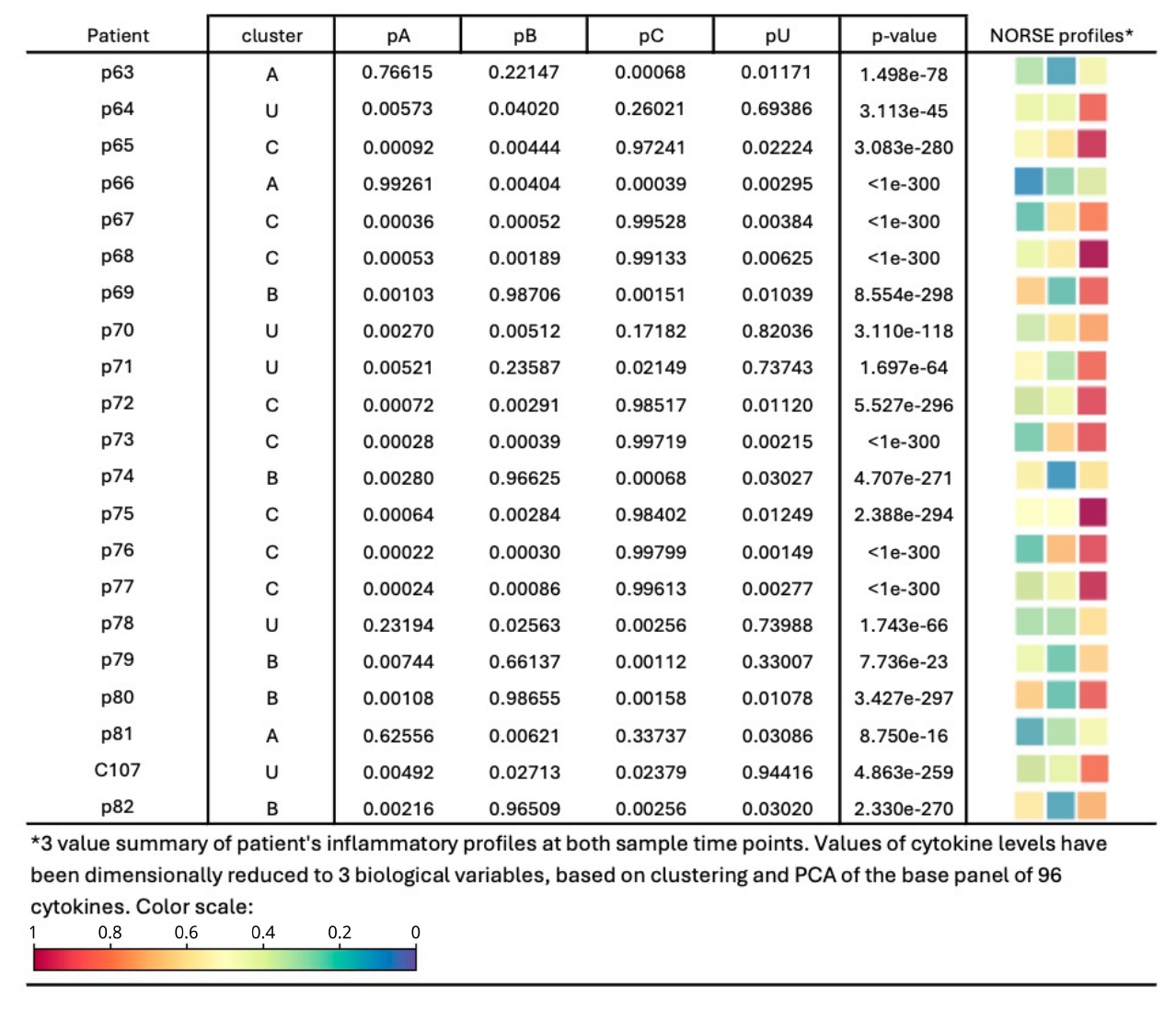}
\caption{Inflammatory profiling summary table for the validation dataset (same biological batch as the NORSELONGI cohort).}
\label{tableS2}
\end{table}

\end{document}